# Generation of robust spatiotemporal optical vortices with transverse orbital angular momentum beyond $10^2$

Wei Chen 1,3*, Wang Zhang 1,3, Yuan Liu 1, Fan-Chao Meng 2, John M. Dudley 2, Yan-Qing Lu 1*

[1] National Laboratory of Solid State Microstructures, Key Laboratory of Intelligent Optical Sensing and Manipulation, College of Engineering and Applied Sciences, and Collaborative Innovation Center of Advanced Microstructures, Nanjing University, Nanjing 210093, China

[2] Institut FEMTO-ST, Université Bourgogne Franche-Comté CNRS UMR 6174, Besançon, 25000, France

[3] These authors contributed equally: Wei Chen, Wang Zhang.

*Correspondence to: wchen@nju.edu.cn, yqlu@nju.edu.cn

**Recently, photons have been observed to possess transverse orbital angular momentum (OAM)[1–5]; however, it is unclear as whether they can hold a transverse OAM higher than 1. Here, we theoretically and experimentally demonstrate that high-order spatiotemporal Bessel optical vortices (STBOVs) can stably carry transverse OAM even beyond $10^2$. Through the inverse design of the spiral phase, an STBOV of any order can be controllably generated using a 4*f* pulse shaper. In contrast to conventional longitudinal OAM, the vector direction of the transverse OAM can be distinguished by the unique time-symmetrical evolution of STBOVs. More interestingly, the stability of STBOVs improves with their increasing orders owing to enhanced space–time coupling[6], making these beams particularly suitable for the generation of ultra-high transverse OAM. Our work paves the way for further research and application of this unique OAM of photons.**

Optical vortices (OVs) are electromagnetic waves that carry the orbital angular momentum (OAM) of photons. OVs with a longitudinal OAM can be generated by introducing an optical spiral phase into the two-dimensional transverse plane (*e.g.*, the *x*–*y* plane) of light fields[7,8], wherein 'longitudinal' means that the OAM vector is parallel to the propagation direction of the beam. In past decades, OVs with a longitudinal OAM have been used in many applications, including optical tweezers[9], super-resolution imaging[10], optical information processing[11], and quantum key distribution[12].

Despite such achievements by OVs with longitudinal OAM, studies have only recently been able to experimentally generate OVs that carry transverse OAM, which is perpendicular to the propagation direction. This transverse OAM beam was first observed in femtosecond filaments in air[1] and was subsequently realized in polychromatic wave packets in free space[2,3], known as spatiotemporal optical vortices (STOVs). The second-harmonic generation of STOVs has been reported very recently[4,5] along with the generation of partially coherent STOVs[13], further verifying that photons can possess intrinsic transverse OAM. However, previous studies on STOV are limited to low transverse OAM, *e.g.*, *l* = 1, 2 (where *l* is the topological charge), in which a higher-order

STOV quickly degrades into several 1-order STOVs during propagation[1-5,13]. This greatly limits the scientific research for and the real-world application of STOVs.

Spatiotemporal astigmatism is one of the reasons contributing to the difficulty in generating a high-order STOV, *i.e.*, in free space, pulsed wave packets generally undergo space diffraction without time diffraction[14]. More importantly, all recent experiments have been realized by directly loading a spiral phase via a conventional pulse shaper[15] that contains a phase device between two gratings in a 4*f* system. In this arrangement, the phase-loaded plane is regarded as the $k_x$–$\omega$ plane and the STOV generally corresponds to a 'patch' on the light-cone[6] (shown in Fig. 1a, where $k_x$ and $k_z$ are the transverse and longitudinal spatial frequencies, respectively, and $\omega$ is the temporal frequency). Because the shaper is only used to realize the temporal modulation, an additional spatial Fourier transform (SFT) is required to convert the spiral phase onto the *x*–*t* plane. This SFT could be realized by adding a cylindrical lens[2] or free-space transmission[3]. Its principle is, essentially, to use the propagation phase to achieve the SFT; therefore, such an approach works well around a certain position—the Fourier plane of the cylindrical lens (or the far field)—and the STOVs (especially for high orders) inevitably separate at other positions (Fig. 1a), which arrests the generation of high-order STOVs in the laboratory. Therefore, there is an urgent need to find a solution for generating stable and high-order STOVs.

Here, we show that according to the inverse design of the spiral phase, a position-sensitive SFT is not required, thereby realizing an STOV without degradation directly from the *x*–$\omega$ phase modulation. We also show that such an operation corresponds to perfect-circle modulation with a spiral phase on the $k_x$–$\omega$ plane (as shown in Fig. 1b), *i.e.*,

$$E_l(k_x, \omega) = \delta\left(\sqrt{k_x^2 + \gamma \Omega^2} - R\right) e^{il\theta_{k_x - \sqrt{\gamma}\Omega}}, \quad (1)$$

where $\Omega$ is the detuning frequency, $\Omega = \omega - \omega_0$, $\omega_0$ is the central frequency, $\gamma$ is the reduction coefficient for temporal and spatial scale consistency, $R$ is the modulated radius, and $\theta_{k_x - \sqrt{\gamma}\Omega}$ is the angular coordinate on the $k_x$–$\omega$ plane, which happens to be the spatiotemporal Bessel optical vortex (STBOV), as follows[16]:

$$E_l(x, z; t) \propto J_l\left(R\sqrt{x^2 + \frac{\tau^2}{\gamma}}\right) \cdot e^{il\theta_{x - \tau/\sqrt{\gamma}}} \cdot e^{i\omega_0(t - z/v_p)}, \quad (2)$$

where $J_l$ is the *l*-order Bessel function of the first kind, $\tau = t - z/v_g$ is the retarded time in the pulse frame where $v_g$ is the group velocity, $v_p$ is the phase velocity, and $\theta_{x - \tau/\sqrt{\gamma}}$ is the angular coordinate on the *x*–$\tau$ plane. We further show that an enhanced space–time coupling[6]—introduced by the one-to-one correspondence between $\omega$ and $k_x$ as described in Eq. (1)—effectively relieves the spatiotemporal astigmatism effect, thereby guaranteeing a robust propagation of STBOVs with $l \geq 10^2$ over 30 Rayleigh distances, which is two orders of magnitude higher than the results reported thus far. Our findings indicate that photons can hold ultra-high transverse OAM for a long time.

To achieve such STBOVs via direct *x*–$\omega$ phase modulation, the spiral phases are inversely designed, inspired by the fact that the space coordinate *x* and the spatial frequency $k_x$ are essentially a Fourier transform pair. That is, one can project a single point in the spatial frequency domain onto a location-shifted grating in the *x* domain, wherein the period and shifted displacement of the grating depend on the amplitude and phase of the frequency point, respectively (Supplementary Fig. S1). The phases for generating STBOVs with topological charges of $l$ = 10, 25, 50, and 100 are shown in Figs. 2a–d, wherein topological charges are manifested as the amount of dislocation between the left and right main lobes in the phase diagram. By applying such phases in the $x$–$\omega$ domain, the inherent one-to-one correspondence between $k_x$ and $\omega$ of the STBOVs will be generated immediately, thereby establishing a compensation mechanism between the space and time diffraction, which ensures that STBOVs have strong energy localisation (Supplementary Information).

We carried out experiments to synthesise STBOVs utilising a custom 4$f$ pulse shaper with a 2D phase-only spatial light modulator (SLM), starting with femtosecond pulses from a Ti:sapphire laser with a central wavelength of ~800 nm. The space–time intensities of STBOVs were measured using a Mach-Zehnder interferometer (Supplementary Fig. S4). Figure 2 shows the experimental and simulated results of STBOVs with topological charges of $l$ = 10, 25, 50, and 100. The measured STBOVs are plotted in Figs. 2e–h, and simulations of the same coordinates are drawn with respective illustrations accordingly. The reconstructed 3D intensity iso-surface profiles of these STBOVs are shown in Supplementary Fig. S5. Notably, the spatial and temporal bandwidths of these STBOVS were set to $\Delta k_x$ = ~123 rad/mm and $\Delta\lambda$ = ~12 nm for temporal and spatial scale consistency. Despite the slight distortion, the experimental results are in good agreement with the simulation results. Besides, the phase reconstruction of high-order STBOVs is difficult due to the air disturbance-induced noise and the limited accuracy of the equipment in the experiment. Nevertheless, the simulated phases of these STBOVs (Figs. 2i–l) verify the spiral phase of the corresponding topological charges, implying the carried transverse OAM (the reconstructed phases for STBOVs with $l$ = 3, 5, 10, and 25 are shown in Supplementary Fig. S6). Furthermore, we calculated the spatial and temporal diameters of the STBOVs from Eq. (2). The theoretical, simulated, and experimental results for STBOV diameters as a function of topological charges are shown in Figs. 2m and 2n, indicating that their diameters are linearly related to the topological charge, except for $l$ = 0.

To investigate the propagation dynamics of STBOVs, we captured two STBOVs with opposite topological charges of $l$ = 100 and -100 at different positions along $z$. Figure 3a shows the intensities of STBOVs with $l$ = 100 at $z$ = -150, -100, -50, 0, 50, 100, and 150 mm. We marked the position of a standard STBOV (described in Eq. (2)) as $z$ = 0 mm and observed a time-symmetrical evolution of the STBOV along the $t$ coordinate. We speculate that the STBOV may carry a certain amount of initial negative chirp, which is gradually compensated by the positive dispersion of air during propagation; as the STBOV gradually accumulates positive chirp, it begins to exhibit a mirror image evolution. The simulated propagation dynamics of the STBOV with $l$ = 100 are shown in Fig. 3c. As expected, by setting a certain amount of initial negative chirp, we obtained simulated results that were consistent with the experimental results. The corresponding simulated phases are shown in Fig. 3d. Interestingly, unlike the longitudinal OAM beam, for which the sign of the topological charge is often difficult to judge by the intensity distribution alone, the results of STBOV with $l$ = -100 are similar to the mirror version of the results with $l$ = 100 (Fig. 3b), which indicates that such transverse OAM may not be simply described using an analogy with the conventional OAM of light. Additionally, we calculated the eccentricity of these two STBOVs at different positions, $\varepsilon = \sqrt{1-(\xi/\eta)^2}$, where $\xi$ and $\eta$ are the major and minor axes of these STBOVs on the $x$-$z$ plane.

The farther the value of $\varepsilon$ is from 0, corresponding to a standard circle, the larger the distortion of the mode is. The evolution of eccentricity further verifies this time-symmetrical evolution of STBOVs, despite the minor deviation between the experiment and the simulation owing to the slight beam divergence caused by incomplete collimation.

We further investigate the propagation dynamics of STBOVs with different topological charges, *i.e.*, $l = 25$ and 0, but at the same positions. For the STBOV with $l = 25$, the time-symmetrical evolution is observed again, as shown in Figs. 4a and 4b. One can clearly see that as the carried transverse OAM decreases, the evolution of the STBOV accelerates. At $z = \pm 150$ mm, the STBOV can no longer maintain the singularity of zero intensity, which is also reflected by the calculated eccentricities. The experimental results of the 0-order STBOV, *i.e.*, the spatiotemporal Bessel beam (STB), reveal such an accelerating effect more vividly (Figs. 4c and 4d). Nevertheless, even for the STB mode, it still maintains approximately seven times the Rayleigh distance of a Gaussian beam with the same full-width at half-maximum (see the time-averaged intensities in Fig. 4e), which is consistent with previous theoretical predictions[16]. For an STBOV with an $l$ greater than $10^2$, its mode can be stably maintained at least 30 times the Rayleigh distance. Compared with the currently reported STOVs, these STBOVs are quite suitable for the generation and carrying of ultra-high transverse OAM, which will benefit further studies on such OAM of light.

We have demonstrated that through the inverse design of the spiral phase, light that carries transverse OAM exceeding $l = 1$ can be controllably generated without degradation. In this work, we demonstrated only the generation of STBOV with $l \leq 10^2$ owing to the size limitation of the SLM in the experiment. Nevertheless, our scheme does not have a theoretical limit (Supplementary Videos 1–4). We note that a few theoretical methods for generating STOVs based on integrated optics have been introduced recently[17,18], while for generating STBOVs with complex frequency relationships, it remains a major challenge to this integrated design. Similar to the conventional OVs with longitudinal OAM, the STBOVs generated using our scheme have the potential to be used to increase the data rate in telecommunications. Owing to the robustness of STBOVs, novel effects, such as 'transverse' spin-orbit angular momentum coupling[19,20] and the evolution of transverse OAM in high harmonic generation[21], may be discovered by further studying the interaction of STBOVs and matter.

In summary, we have theoretically and experimentally demonstrated the generation of high-order STBOVs with transverse OAM, even beyond $10^2$. To the best of our knowledge, this is two orders of magnitude higher than the existing results. More importantly, this work not only provides the first evidence that photons can hold ultra-high transverse OAM but also proposes a universal strategy with high operability and flexibility to generate such OAM. By investigating the propagation dynamics of STBOVs, we found that STBOVs with initial negative chirps exhibit a time-symmetrical evolution when propagating in air. Theoretical and experimental studies have also indicated that with an increase in $l$, the evolution of STBOV decelerates, which is conducive to the further generation and study of higher-order STBOVs. Furthermore, our method is also applicable to other wave systems, such as material waves, water waves, acoustic waves, microwaves, and surface plasmon polaritons, and this opens the door for transverse OAM applications.

## Data availability

The data that support the plots within this paper and other findings of this study are available from the corresponding authors upon reasonable request.

# Acknowledgements

The work was supported by the National Key Research and Development Program of China (2017YFA0303700) and the Contract ANR-15-IDEX-0003 and ANR-17-EURE-0002.

# Author contributions

# Competing interests

The authors declare no competing interests

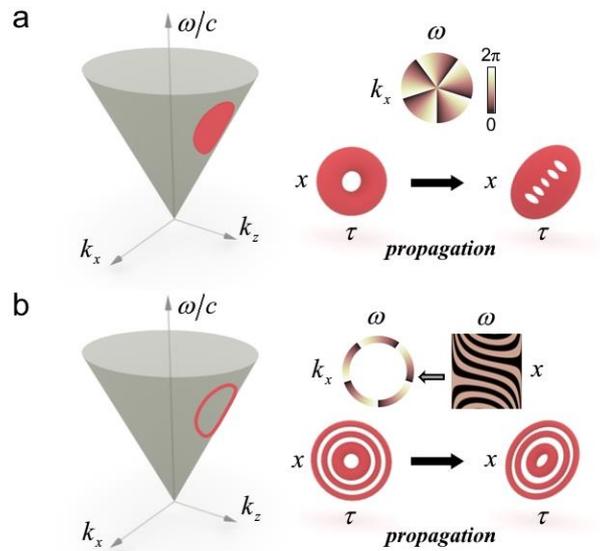

**Figure 1| Concept of an STBOV. a**, An STOV with $l = 5$ corresponds to a 'patch' with spiral phase on the light-cone $k_x^2 + k_z^2 = (\omega/c)^2$. Notably, the current scheme cannot effectively produce such a spiral phase on the $k_x$–$\omega$ plane, and the 5-order STOV degrades into five 1-order STOVs rapidly owing to its reliance on position-sensitive SFT. **b**, Inverse design of the phase makes it possible to effectively produce perfect-circle modulation with a spiral phase on the $k_x$–$\omega$ plane, leading to the generation of a robust 5-order STBOV.

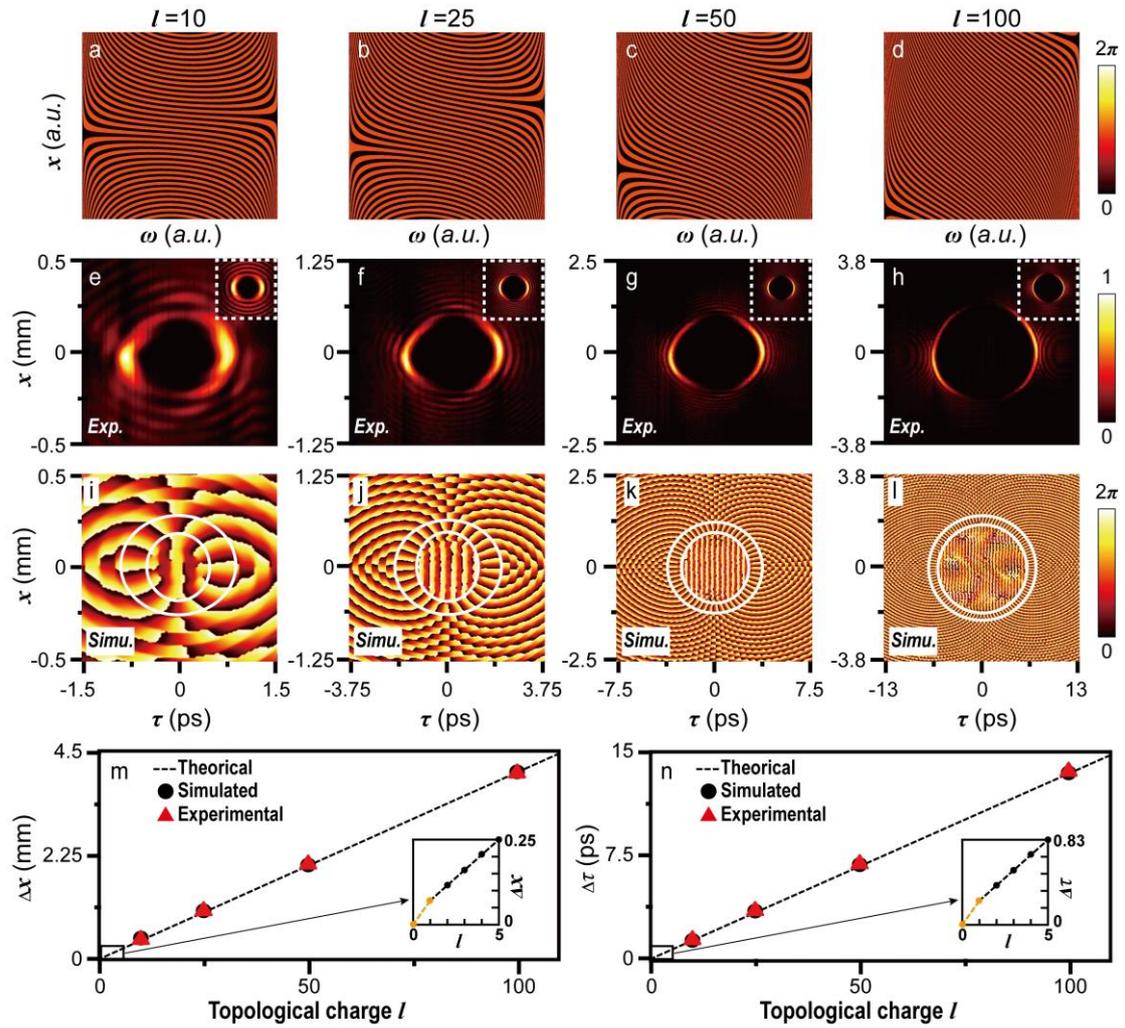

**Figure 2 | Experimental and simulated results of STBOVs with topological charges of $l$ = 10, 25, 50, and 100:** SLM phase patterns (**a–d**), measured space–time intensities (**e–h**), simulated space–time phases (**i–l**), and the spatial or temporal diameter dependence (**m**, **n**) of such STBOVs. The white dotted squares (circles) in **e–h** (**i–l**) show the simulated intensities (spiral phases) of each STBOV. The corresponding topological charges are marked at the top of each column.

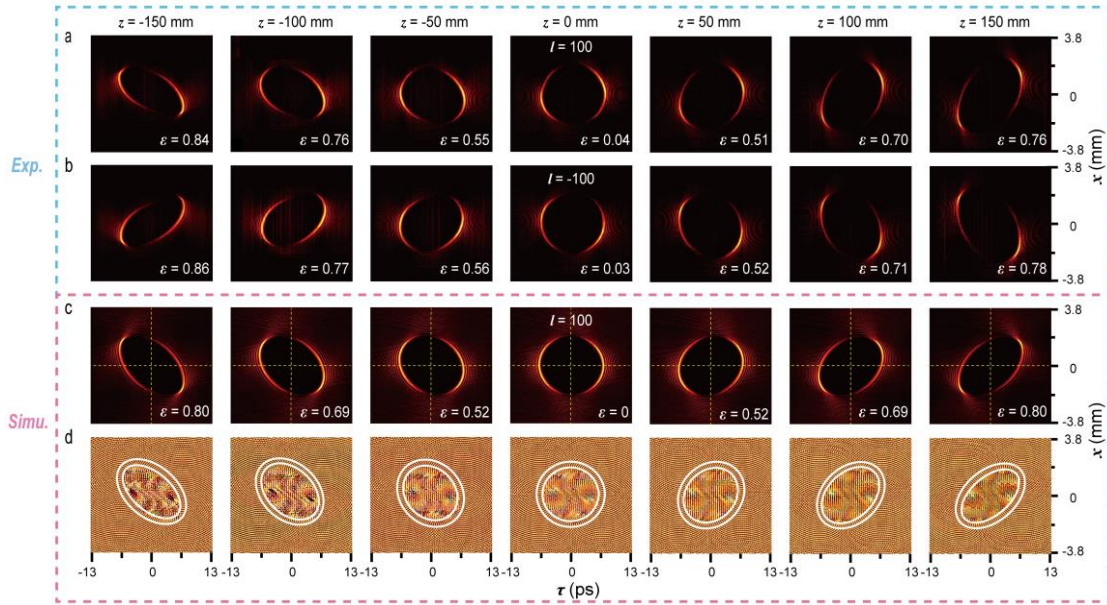

**Figure 3 | Experimental and simulated results of the propagation dynamics of two STBOVs with topological charges of $l$ = 100 and -100. a**, Measured intensities of an STBOV with the topological charge of $l$ = 100 at $z$ = -150, -100, -50, 0, 50, 100, and 150 mm, where the position of the standard STBOV (described in Eq. (2)) is marked as $z$ = 0 mm. **b**, Same as **a**, but with a topological charge of $l$ = -100. **c**, **d**, Simulated intensities (upper panels) and phases (lower panels) for $l$ = 100. The corresponding eccentricities are given in the lower right corner of each figure in **a–c**. The positions are marked at the top of each column.

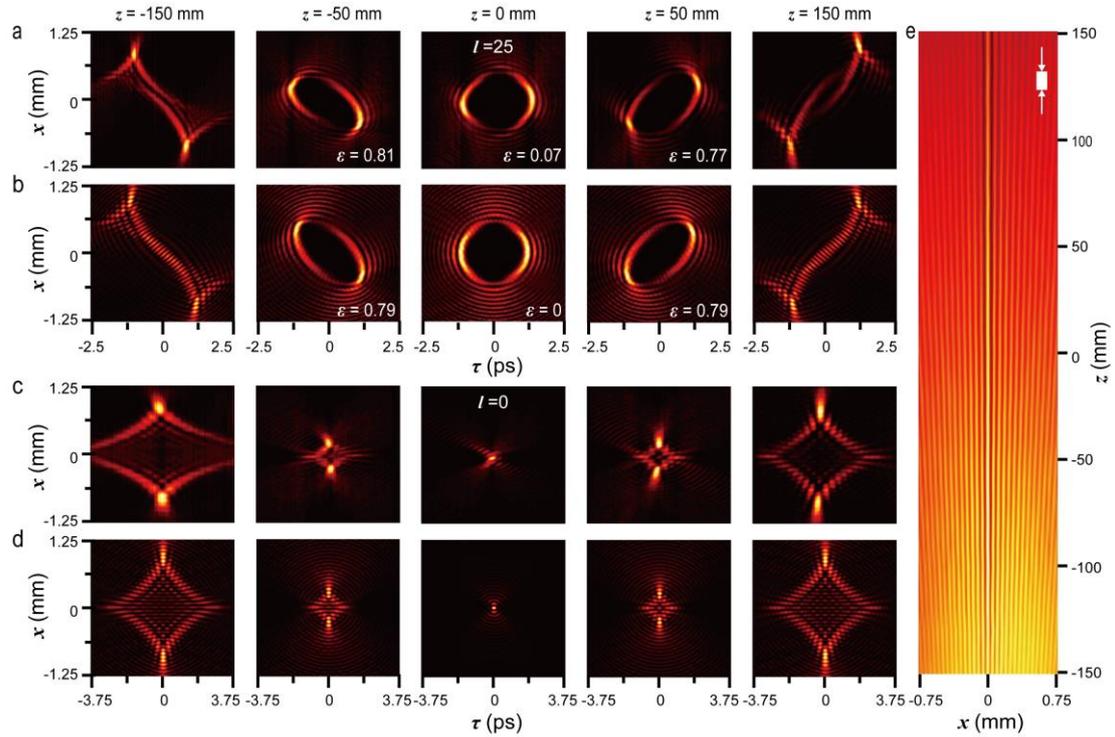

**Figure 4 | Experimental and simulated results of propagation dynamics of two STBOVs with topological charges of $l$ = 25 and 0. a**, Measured intensities of an STBOV with the topological charge of $l$ = 25 at $z$ = -150, -50, 0, 50, and 150 mm, where the position of the standard STBOV (described in Eq. (1)) is marked as $z$ = 0 mm. **b**, Corresponding simulated results of **a**. **c**, Same as **a**, but with the topological charge of $l$ = 0. **d**, Corresponding simulated results of **c**. The positions are marked at the top of each column in **a–d**. **e**, Measured time-averaged intensities of the STBOV with $l$ = 0 along $z$, where the solid white rectangle in the top-right corner represents the Rayleigh distance $Z_R$ = 8.6 mm for the Gaussian beam with $\Delta x$ = ~55 μm. The time-averaged intensities at different positions also reflect the slight divergence of the light beam in the experiment.

# Supplementary information:
# Generation of robust spatiotemporal optical vortices with transverse orbital angular momentum beyond $10^2$

## I. Supplementary Theory

### 1) Principle of inverse design of spiral phase

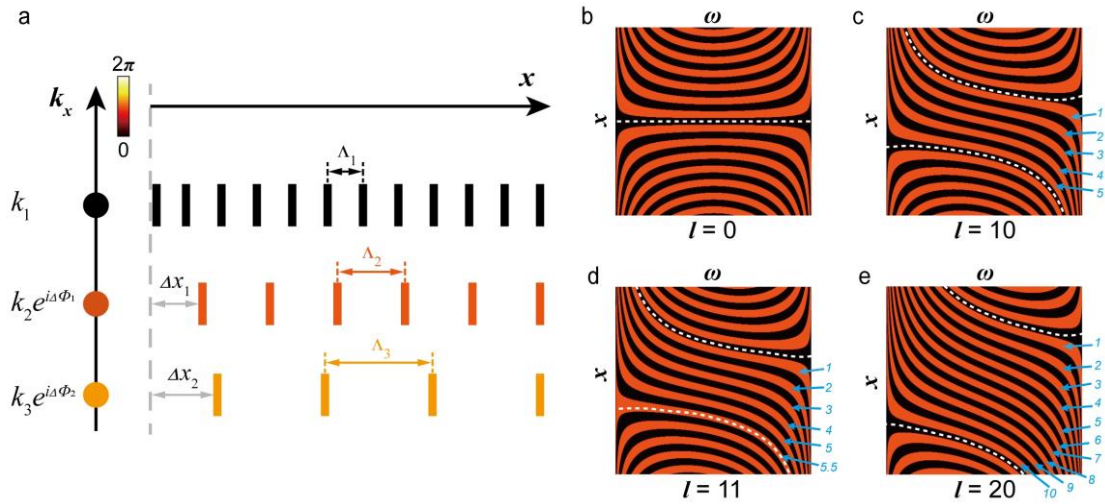

**Figure S1 | Principle of inverse design of phase. a**, Principle. **b–e**, Phase patterns for generating STBOVs with different topological charges of $l = 0, 10, 11$ and $20$. The white dashed lines in **b–e** mark the positions of the left and right main lobes, and the blue arrows point out the accumulated dislocation of phase between the left and right main lobes.

As shown in Fig. S1a, for a frequency point $E(\omega_0, k_n)\delta(k_x - k_n)$ (with phase factor $e^{i\phi_n}$) lying on the spatial frequency $k_x$ axis, its counterpart on the real $x$ axis is actually a location-shifted grating, with a period of $\Lambda_n = 2\pi/k_n$, and shifted displacement $\Delta x_n$ of the grating is proportional to the phase of the frequency point, $\Delta x_n \propto \phi_n$. Hence, according to this principle, we can obtain the $x$–$\omega$ phase of a STBOV that possesses a perfect-circle-modulation with a spiral phase on its $k_x$–$\omega$ plane. For an STBOV, the amount of dislocation between the left and right main lobes in the $x$–$\omega$ phase diagram reflects its topological charge, $i.e.$, the dislocation of phase between the left and right main lobes can be described as $(l/2)\times\pi$. As shown in Figs. S1b–e, for STBOVs with topological charges of $l = 0, 10, 11$, and $20$, the corresponding dislocations are $0, 5\pi, 5.5\pi$, and $10\pi$. Notably, the phase binarization brings additional high-frequency components, which can be removed by spatial filtering in the experiment.

## 2) Enhanced space–time coupling in STBOVs

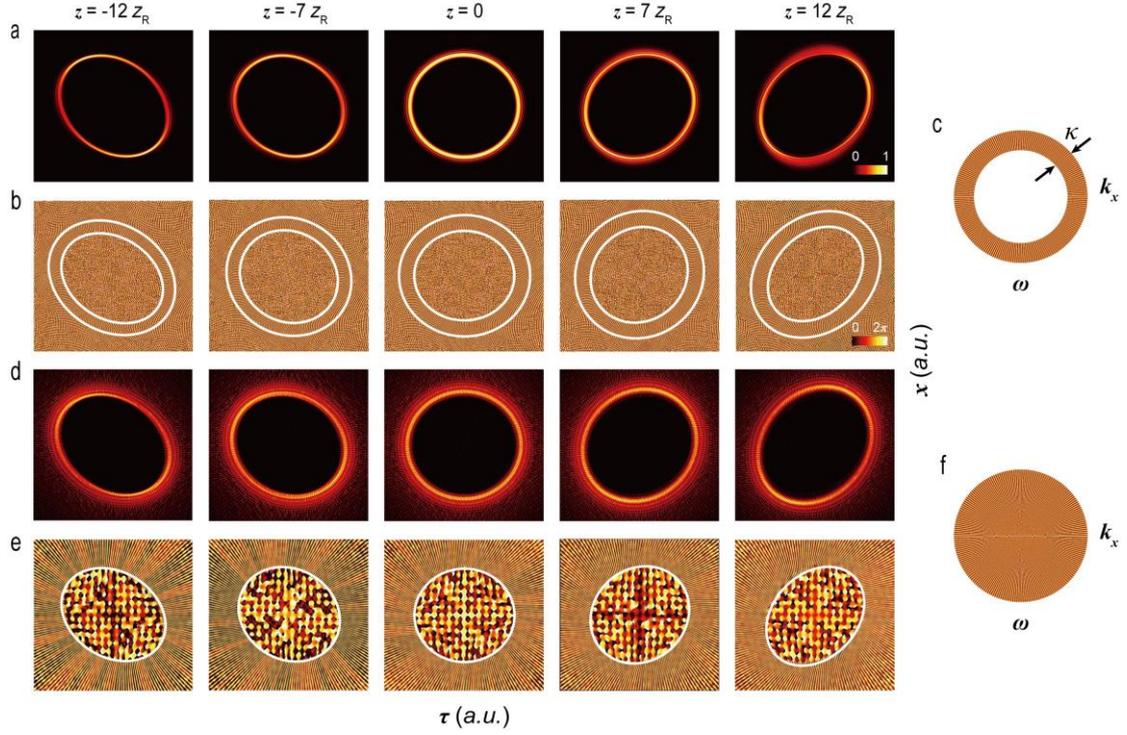

**Figure S2 | Simulated results of propagation dynamics of two STBOVs with the same topological charge of $l = 200$ but different spectral uncertainties $\kappa$. a**, Simulated intensities of an STBOV with low spectral uncertainty, where the position of the standard STBOV is marked as $z = 0$. **b**, Corresponding phases of **a**. **c**, Corresponding spatiotemporal spectra of **a**. **d**–**f**, same as **a**–**c**, but with high spectral uncertainty. The white dotted circles in **b** and **e** mark the spiral phases of each STBOV. The positions are marked at the top of each column (of **a**, **b**, **d**, and **e**), where $Z_R$ is the Rayleigh distance of a Gaussian beam with the same full-width at half-maximum of the corresponding 0-order STBOV.

As shown in Fig. S2a, the energy of an STBOV remains non-spreading during propagation. This is because the spatiotemporal spectra of the STBOV (Fig. S2c) have low spectral uncertainty $\kappa$, that is, $k_x$ and $\omega$ correspond one to one. Such one-to-one correspondence enhances space–time coupling[1] and establishes a compensation mechanism between the space and time diffraction, thereby effectively relieving the spatiotemporal astigmatism effect. As $\kappa$ increases, the compensation mechanism gradually fails, causing the energy of the beam to rapidly diffuse toward the edge, as shown in Figs. S2d–f.

# 3) Theoretical analysis of the equivalence of the $x$–$\omega$ modulation scheme

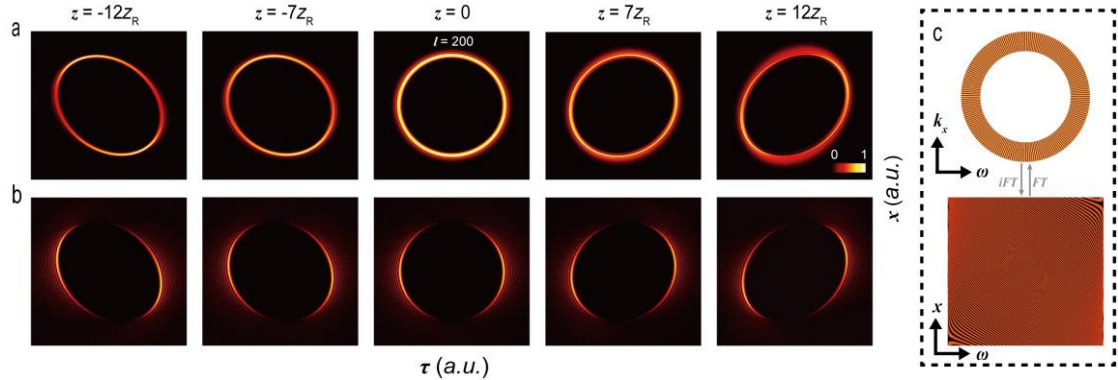

**Figure S3 | Simulated results of propagation dynamics of two STBOVs with the same topological charge of $l$ = 200 but undergoing different spatiotemporal spectra modulation. a**, Simulated intensities of an STBOV through $k_x$–$\omega$ modulation. **b**, Simulated intensities of a STBOV through $x$–$\omega$ modulation. **c**, Phase relationship between $k_x$–$\omega$ and $x$–$\omega$ modulation. The positions are marked at the top of each column (of **a** and **b**), where $Z_R$ is the Rayleigh distance of a Gaussian beam with the same full-width at half-maximum of the corresponding 0-order STBOV.

As shown in Fig. S3a, the ideal STBOV, represented as a perfect circle with a spiral phase on the $k_x$–$\omega$ plane, has a uniform mode distribution of energy. However, as we analysed in the main text, it is difficult to directly perform such $k_x$–$\omega$ modulation in experiments because of the indispensable position-sensitive spatial Fourier transform (SFT). Fortunately, regardless of the slight loss of energy in the $x$ direction away from the centre (owing to the pixel accuracy), the $x$–$\omega$ modulation scheme proposed in this work can be an effective method to generate STBOVs, as shown in Fig. S3b.

## II. Supplementary Experimental Setup

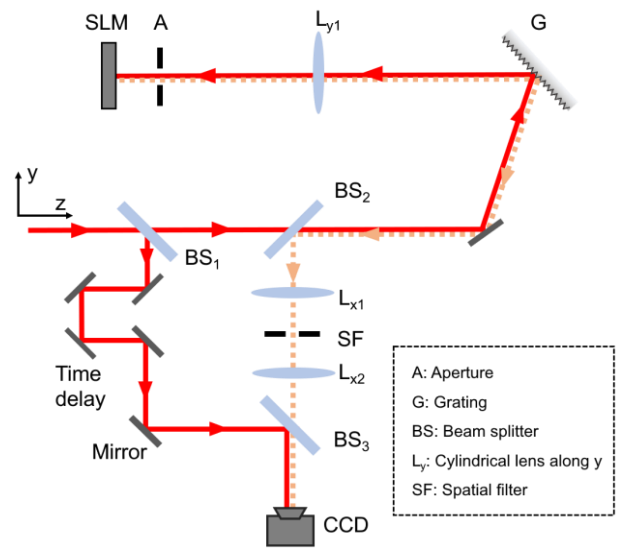

**Figure S4 | Experimental setup, which consists of two sections, for generating and characterizing STBOVs:** (1) STBOV generator consisting of a grating, a cylindrical lens $Ly_1$, an aperture, and an SLM; and (2) a time-resolved profile analyser that is realized by a Mach–Zehnder interferometer consisting of two BSes ($BS_1$ and $BS_3$), a CCD camera, and a motorized translation stage in the reference path. L: cylindrical lens, SLM: spatial light modulator, CCD: charge-coupled device, BS: beam splitter.

To generate STBOVs, a 4*f* pulse shaper consisting of a diffraction grating, a cylindrical lens, and an SLM was used, as shown in Fig. S4. The frequencies of an ultrashort pulse are spatially spread by the grating and collimated to the SLM via a cylindrical lens, which can be understood as a temporal Fourier transform. We consider the SLM to be the *x*–*ω* plane, where the phase patterns for the generation of STBOVs are loaded. After the SLM, the light field is retroreflected and reconstituted at the same diffraction grating, thereby immediately generating the STBOVs, with no need for an SFT. To extract the space–time intensities of STBOVs, a Mach–Zehnder interferometer was used in a similar way as those described by Li and Lewellen[2] and Li *et al.*[3].

## III. 3D Iso-Intensity Profiles of STBOVs

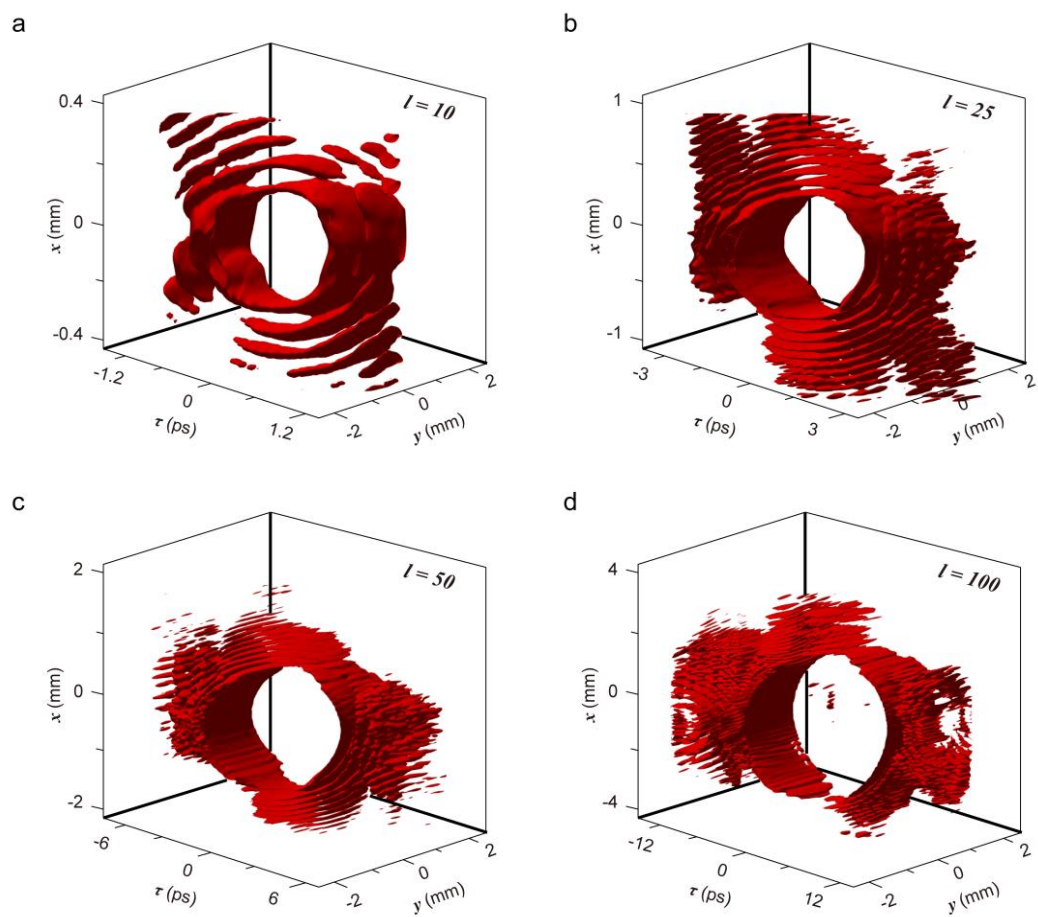

**Figure S5 | Measured 3D profiles of STBOVs with *l* = 10, 25, 50, and 100.** The method for 3D iso-intensity reconstruction is according to Ref. 2 and 3.

# IV. Phase Reconstruction of STBOVs

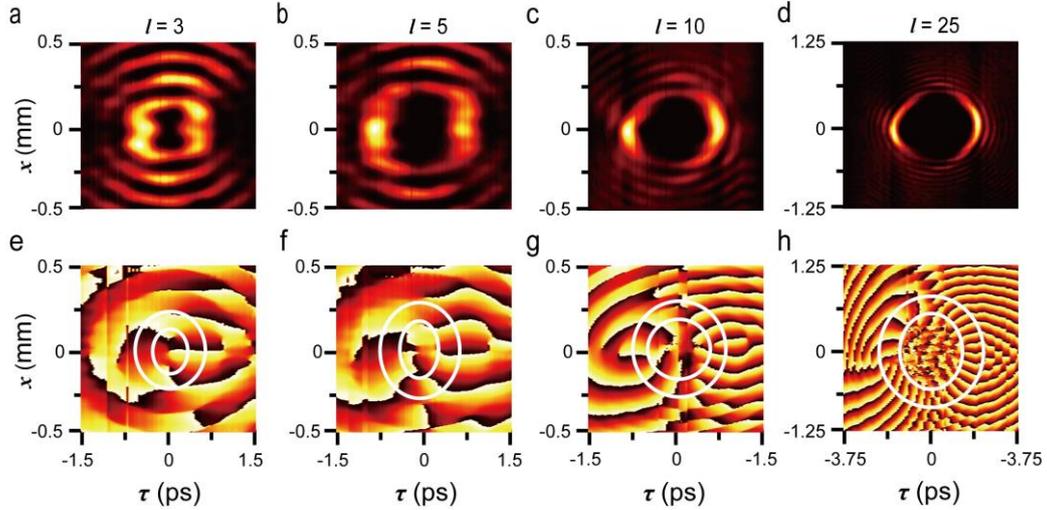

**Figure S6 | Measured intensities and reconstructed phases of STBOVs with *l* = 3, 5, 10, and 25. a–b**, Measured intensities of STBOVs. **e–h**, Reconstructed phases of STBOVs accordingly. The white circles in **e–h** mark the spiral phases of each STBOV. The corresponding topological charges are marked at the top of each column.

The space–time intensities of STBOVs with $l$ = 3, 5, 10, and 25 are shown in Figs. S6a–d. As shown in Figs. S6e–h, the corresponding reconstructed phases verify the spiral phase of corresponding topological charges despite the slight distortion, implying the carried transverse OAM. Notably, the spatial bandwidth and temporal bandwidth of these STBOVS are set to $\Delta k_x$ = ~61.5 rad/mm and $\Delta \lambda$ = ~6 nm for $l$ = 3 and 5.

# V. Supplementary Data

Video S1. Theoretical movie for propagation dynamics of a STBOV with $l$ = 200 generated by the $k_x$–$\omega$ modulation.

Video S2. Theoretical movie for propagation dynamics of the same STBOV generated by the $x$–$\omega$ modulation.

Video S3. Theoretical movie for propagation dynamics of a STBOV with $l$ = 1000 generated by the $x$–$\omega$ modulation.

Video S4. Theoretical movie for propagation dynamics of a STBOV with $l$ = -1000 generated by the $x$–$\omega$ modulation.